\documentclass[numberedappendix,12pt,preprint]{emulateapj}

\usepackage[english]{babel}
\usepackage{amsmath,amsfonts,amsthm}
\usepackage{amssymb}
\usepackage{graphicx}
\usepackage[pdftex, colorlinks=true, citecolor=blue, linkcolor=blue]{hyperref}

\begin{document}
\title{The Cosmic Web Around The Brightest Galaxies During The Epoch Of Reionization}

\author{Keven Ren\altaffiliation{1,2,$\dagger$}, Michele Trenti\altaffiliation{1,2}, and Simon Mutch\altaffiliation{1,2}}

\affil{$^1$ School of Physics, The University of Melbourne, Parkville, Victoria, Australia}
\affil{$^2$ ARC Centre of Excellence for All Sky Astrophysics in 3 Dimensions (ASTRO 3D)}

\email{$^{\dagger}$ kevenr@student.unimelb.edu.au; mtrenti@unimelb.edu.au}

\begin{abstract}

The most luminous galaxies at high-redshift are generally considered to be hosted in massive dark-matter halos of comparable number density, hence residing at the center of overdensities/protoclusters. We assess the validity of this assumption by investigating the clustering around the brightest galaxies populating the cosmic web at redshift $z\sim8-9$ through a combination of semi-analytic modeling and Monte Carlo simulations of mock Hubble Space Telescope WFC3 observations. The innovative aspect of our approach is the inclusion of a log-normal scatter parameter $\Sigma$ in the galaxy luminosity versus halo mass relation, extending to high-$z$ the conditional luminosity function framework extensively used at low redshift. Our analysis shows that the larger the value of $\Sigma$, the less likely that the brightest source in a given volume is hosted in the most massive halo, and hence the weaker the overdensity of neighbors. We derive a minimum value of $\Sigma$ as a function of redshift by considering stochasticity in the halo assembly times, which affects galaxy ages and star formation rates in our modeling. We show that $\Sigma_{min}(z)\sim0.15-0.3$, with $\Sigma_{min}$ increasing with redshift as a consequence of shorter halo assembly periods at higher redshifts. Current observations ($m_{AB}\sim27$) of the environment of spectroscopically confirmed bright sources at $z>7.5$ do not show strong evidence of clustering and are consistent with our modeling predictions for $\Sigma\geq\Sigma_{min}$. Deeper future observations reaching $m_{AB}\sim28.2-29$ would have the opportunity to clearly quantify the clustering strength, and hence to constrain $\Sigma$, investigating the physical processes that drive star formation in the early Universe. 
\end{abstract}

\section{INTRODUCTION \label{sec:intro}}

The transition from a homogeneous universe to a structured one is characterized in the Cold Dark Matter (CDM) paradigm by the gradual hierarchical merging of smaller haloes and gas into larger structures over time \citep{1985ApJ...292..371D}. Thus the study of early-time objects is crucial in understanding the evolution of the Universe from its initial stages of structure formation during the Epoch of Reionization, a period where the radiation of the first stars begin to reionize the clouds of neutral hydrogen permeating the Universe, to a state that is rich in structural features endemic of its current climate.  As such, modern observational surveys are constantly probing at higher redshifts in an effort to locate and characterize these first structures; rare early quasars via the Sloan Digital Sky Survey (SDSS) \citep{2002AJ....123.2945R}, and galaxies through the Hubble Deep Field \citep{1996AJ....112.1335W} and Ultra-Deep Field \citep{2006AJ....132.1729B, 2013ApJS..209....6I} campaigns. The latter (small-area) programs are effectively complemented by large area surveys such as the Cosmic Assembly Near-infrared Deep Extragalactic Legacy Survey (CANDELS) \citep{2011ApJS..197...35G, 2011ApJS..197...36K}, the Brightest of Reionizing Galaxies (BoRG) survey \citep{2011ApJ...727L..39T, 2016ApJ...817..120C}, and the \emph{Frontier Fields} \citep{2017ApJ...837...97L}, which are identifying objects from within the first billion years after the Big Bang (e.g. \citealt{2014ApJ...793L..12Z, 2015ApJ...810L..12Z, 2015ApJ...804L..30O, 2016ApJ...819..129O, 2016ApJ...827...76B,2017ApJ...835..113L}). Still, searching for these distant galaxies and quasars using broad wavelength filters and spectroscopically confirming candidates remain a challenge, more-so for ground-based observations, as the Lyman Break typically shifts outside of the optical bands into the near-IR regime for objects beyond $z \sim 6.5$. As a result, to date there have been spectroscopic confirmations of only a handful of galaxies at redshifts above 7 (with ages less than 750Myr), despite having samples of $\sim 1000$ photometric candidates \citep{2015ApJ...803...34B, 2015ApJ...800...84C, 2017arXiv170604613S}.

Intuitively, the first structures are expected to lie inside regions where the fluctuations in the matter density field are the highest. Furthermore, objects that were formed in these regions are expected to be massive and increasingly clustered by similarly sized objects \citep{1984ApJ...284L...9K, 1988MNRAS.230P...5E}. This characteristic allows for a tentative assessment of an object's high redshift identity to be made given it's mass and expected number of neighbors \citep{2008MNRAS.385.2175M, 2010ApJ...716L.229R}. The recent spectroscopic confirmation of galaxies EGS-zs8-1 \citep{2015ApJ...804L..30O} and EGSY8p7 \citep{2015ApJ...810L..12Z} at $z = 7.73$ and $z= 8.68$ present an ideal testbed to investigate the relation between brightness/mass and clustering properties. In fact, these two sources have exceptionally bright magnitudes ($M_{UV} \approx -22$, to be compared with the characteristic magnitude $M^{*}_{UV} \approx -20.2$ at $z \sim 8$  \citealt{2015ApJ...803...34B}). Consequently, these galaxies are objects with number densities of the order $\sim 10^{-6}$Mpc$^{-3}$ \citep{2015ApJ...803...34B}, which are rare in cosmological simulations of volumes comparable to (or a few times larger than) that of the CANDELS survey \citep{2016MNRAS.461L..51W}. In turn, the luminosity and rarity of the observed galaxies would suggest that these objects live inside massive dark matter haloes, and therefore one would expect to observe neighboring galaxies in its vicinity \citep{2008MNRAS.385.2175M, 2012ApJ...746...55T}. Curiously, neither EGS-zs8-1 or EGSY8p7 show evidence of being located in significantly clustered regions based on current photometric data in the EGS fields.

Moreover, this situation is paralleled by observations of QSO's at $z \gtrsim 6$. These objects are often assumed to reside in extremely massive haloes for their epoch (halo mass $M_h\sim 10^{12} - 10^{13} M_{\odot}$ at $z\sim6$; \citealp{2005Natur.435..629S}), and modeling predicts an excess of surrounding dark-matter halos that should host galaxies detectable by Hubble \citep{2009MNRAS.394..577O, 2011ApJ...736...66R}. Yet, observations of their environments do not identify a clear excess of nearby galaxies compared to random fields \citep{2009ApJ...695..809K, 2014AJ....148...73M}. 

In this work we propose an extension of the typical abundance matching modeling used to simulate clustering around such sources to see whether the apparent mismatch between data and model can be alleviated. Specifically we investigate the impact of going beyond a 1-1 relation between galaxy light and dark matter halo mass adopted by the studies above, and model the inherent stochasticity present in the dark matter halo mass to galaxy luminosity relations.

One way to accomplish this is the use of a statistical approach to link galaxy luminosities to dark matter halo masses to create a distribution of the galaxy luminosity given its host halo mass. This idea has been developed to model galaxies at lower redshift and is called the conditional luminosity function (CLF) approach (see \citealt{2003MNRAS.339.1057Y, 2004MNRAS.353..189V, 2005ApJ...627L..89C}). The dispersion in galaxy luminosities is characterized by a parameter, $\Sigma$ that is assumed log-normal \citep{2005MNRAS.358..217Y}, as this functional form naturally provides an explanation to the observed scatter in the \citet{2000ApJ...533..744T} relation.

Our approach combines the CLF (addressing galaxy to galaxy variations in UV luminosity) with a model capable of predicting galaxy luminosity versus halo mass relations at higher redshifts \citep{2015ApJ...813...21M}, and with dark-matter halo catalogs from cosmological simulations to obtain predictions for the spatial distribution of galaxies around massive/luminous objects at $z\gtrsim 6$. 

This paper is organized as follows. In Section~\ref{sec:method} we describe the model and method used. In Section~\ref{sec:cons} we derive the lower limit of the log-normal scatter amplitude in the CLF using the distribution of the halo assembly times, and compare such scatter against observational data and results from cosmological simulations. In Section~\ref{sec:results} we present our results on the clustering around objects such as EGS-zs8-1 and EGSY8p7. We conclude in Section~\ref{sec:disc}, summarizing our findings. Modeling and simulations in this work use the cosmological parameters from \citet{2016A&A...594A..13P} with $\Omega_{m} = 0.308, \Omega_{b} =0.0484, \Omega_{\Lambda}=0.692, h=0.678, \sigma_{8}=0.815, n_{s}=0.968$. Magnitudes are given in the AB system \citep{1983ApJ...266..713O}.

\section{MODEL DESCRIPTION \& METHOD}  \label{sec:method}

To investigate the clustering properties of rare, very luminous ($L>L_*$) galaxies we turn to Monte Carlo modeling to assign galaxy luminosities to dark matter halos identified in cosmological dark-matter only simulations, using the conditional luminosity function (CLF) method \citep{2005ApJ...627L..89C}. The CLF approach requires to define both an average galaxy luminosity for a given halo mass, that we derive under the semi-analytic framework by \citet{2015ApJ...813...21M}, and a scatter parameter, $\Sigma$, which represents the log-normal dispersion in galaxy luminosities (a free parameter in our model). Our Monte Carlo simulations then stochastically assign a galaxy luminosity to each dark matter halo, drawing pencil beams around the brightest objects to recreate mock catalogs similar to the Hubble's WFC3 field of views for galaxies EGS-zs8-1 and EGSY8p7. We vary the amount of luminosity scatter, $\Sigma$ to assess how this affects the distribution of number counts for galaxies close in projection (and at similar redshift) to these bright high-redshift sources.

\subsection{Halo Catalogues}

\begin{figure}[ht!!]
	\centerline{\includegraphics[trim={0.5cm 0.6cm 0 0.5cm}, clip, angle=-00, scale=0.80]{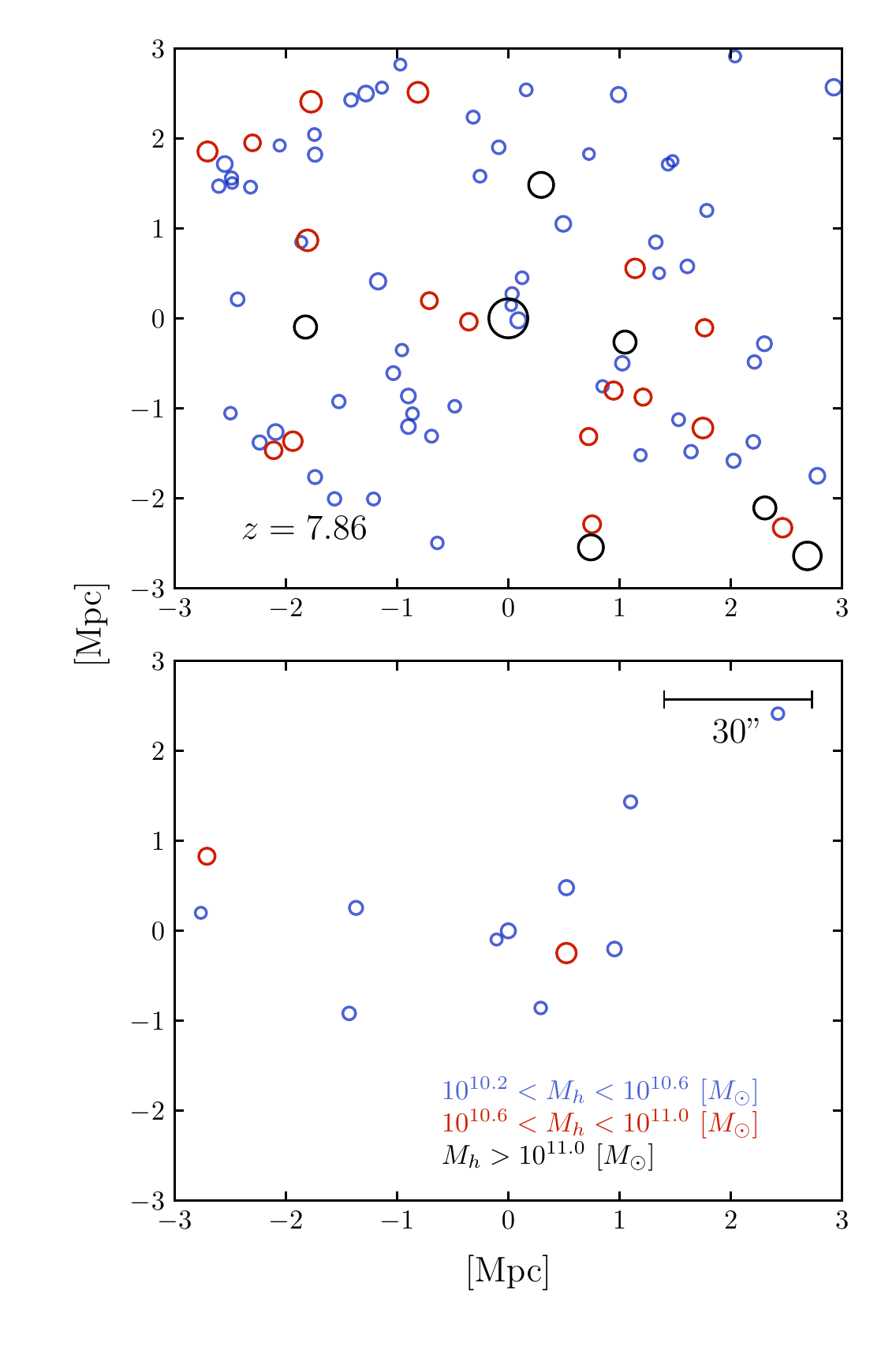}}
	\caption{\small Projected positions of haloes along a simulated pencil beam with a depth of $50$Mpc (comoving) at redshift $z=7.86$. The relative sizes of the markers are indicative of their halo mass. (Upper panel) Pencil beam centered around one of the most massive dark matter haloes in the simulation, with mass $M_{h} \sim 10^{12} M_{\odot}$. (Lower panel) Pencil beam centered across a smaller halo of mass $M_{h} \sim 10^{10.5} M_{\odot}$. Prominent clustering characteristics are a typical feature seen around massive halos compared to lower mass counterparts.}
	\label{fig:beam_compare}
\end{figure}

As a base for our Monte Carlo simulations we utilize dark halo catalogues at redshifts $z=7.86$ and $z=8.64$ derived from the collisionless N-body run, $\tt{Tiamat}$, of the Dark-ages Reionization and Galaxy-formation Observables from Numerical Simulations (DRAGONS) project \citep{2016MNRAS.459.3025P}. $\tt{Tiamat}$ was run with the $\tt{GADGET-2}$ code \citep{2005MNRAS.364.1105S} and uses the $\tt{SUBFIND}$ \citep{2001MNRAS.328..726S} algorithm to identify haloes. $\tt{Tiamat}$ employs a standard cosmology as determined by the \citet{2016A&A...594A..13P}, and was initialized with $2160^{3}$ particles, each with mass $3.89 \times 10^{6} M_{\odot}$. The dark matter halo catalogue we consider extends over a mass range of $10^{9}-10^{12} M_{\odot}$ for $z=7.86$ and $10^{9} - 10^{11.8}M_{\odot}$ for $z=8.64$. The total volume is a cube with edge length of $100$ Mpc comoving with periodic boundary conditions. 

\subsection{Pencil Beam Construction}

We construct a parallelepiped pencil beam through the simulation volume with a depth of $\Delta z = 0.9$, equivalent to $273$ Mpc at $z = 7.86$ and $214$ Mpc at $z=8.68$. This is consistent with the typical uncertainty in photometric redshift for galaxy surveys at high redshift \citep{2015ApJ...803...34B}. For the field of view area (cross sectional slice), the pencil beam is taken to be $2.2' \times 2.2'$ ($\sim 5.9\times 5.9$ Mpc$^{2}$ at both redshifts), approximately matching the field of view of a single pointing of Hubble's Wide Field Camera 3 (WFC3), which is $2.1' \times 2.3'$. To a first approximation, the field of view area is constant along the line of sight, as the angular diameter is relatively insensitive to redshift changes at $z \gtrsim 3$ \citep{1999astro.ph..5116H}. Pencil beams are initialized either around a targeted (bright) object or at random positions (to set a baseline for comparison), and traverse multiple replications of the cube because of periodic boundary conditions. The orientation of the pencil beam are randomly selected around two distinct axes, with their two rotation angles uniformly sampled between $\theta,\phi \in [\pi/6,5\pi/12]$. We limit the range of allowed angles to minimize the incidence of overlapping the beam onto itself (see \citealt{2007MNRAS.376....2K,2008ApJ...676..767T}). Any Monte Carlo run with pencil beam overlap (incidence rate of $\sim 0.5\%$) is removed from the analysis and a replacement with no overlap is generated instead.

\subsection{Conditional Luminosity Function} \label{subsec:cllf}

We use a CLF approach following \citet{2005ApJ...627L..89C} to ``paint'' luminous galaxies onto dark matter haloes. The CLF, $\Phi(L \mid M_{h})$, gives the probability of observing a galaxy of UV luminosity $L$ given a halo mass $M_h$, and has a log-normal form in $L$:

\begin{equation}
\Phi(L \mid M_h)=\dfrac{1}{\sqrt{2\pi}\Sigma L}\exp{\bigg( -\dfrac{\log \Big[ {\frac{L}{\langle L^{*}(M_{h}, \Sigma,z)\rangle}} \Big]^{2}}{2\Sigma^{2}} \bigg) },
\label{eqn:clf}
\end{equation}

where $\Sigma$ is the dispersion (a free parameter in our model), and $\langle L^{*}(M_{h}, \Sigma,z)\rangle$ is the shifted average galaxy luminosity at redshift $z$ (derived in Section \ref{subsec:md}) below. 

\subsection{Galaxy Luminosity Versus Halo Mass Relation} \label{subsec:md}

$\langle L^{*}(M_{h}, \Sigma,z)\rangle$ is a rescaled version of the average galaxy luminosity versus halo mass relation, $L(M_{h},z)$, with the latter based on the semi-analytic framework developed by \citet{2010ApJ...714L.202T}, \citet{2013ApJ...768L..37T} and \citet{2015ApJ...813...21M}. The model is based on the following key assumptions: (1) The predominant factor behind the evolution of the galaxy LF is the assembly of their host dark matter haloes; (2) Stars are formed with a characteristic time given by the median of the dark-matter halo assembly time distribution \citep{1993MNRAS.262..627L,2007MNRAS.376..977G}; (3) The star formation efficiency ($\varepsilon(M_{h})$ defined as the ratio between the stellar mass over halo mass) is dependent only on the halo mass, $M_{h}$, but not on redshift; (4) Stellar luminosities are defined by Single Stellar Populations (SSP) models \citep{2003MNRAS.344.1000B}. A correction in UV luminosities from interstellar dust attenuation is taken into account in the modeling to ensure self-consistency with Hubble observations, using the UV continuum slope measurements from \citet{2014ApJ...793..115B} and following the same approach adopted by \citet{2015ApJ...802..103T} and \citet{2015ApJ...813...21M}. Finally, $\varepsilon(M_{h})$ is derived by calibration at a single redshift through abundance matching between a well constrained observed galaxy UV luminosity function (such as $z \sim 5$) and the theoretical halo mass function. 

These assumptions provide the general implication that the final stellar masses of simulated galaxies depend only on halo mass, with both younger stellar ages and an increasing star formation rates at higher redshifts for a fixed halo mass (as the time required to assemble halos are shorter at higher redshifts). Since the greatest contribution to the galaxy UV luminosity comes from stars younger than $\sim100$Myr \citep{2014ARA&A..52..415M}, the inclusion of multiple epochs of star formation during a halo assembly history improves modeling star-forming galaxies at high redshifts ($z \gtrsim 8$) when the typical assembly period is shorter than $<100$Myr, and ensures self-consistency of the framework as well. 
Here, we consider a two-step model (sufficient for accurate modeling; see \citealt{2015ApJ...813...21M}) so that for a galaxy hosted in a halo of mass $M_h$ at redshift $z$, the star formation history is modeled as two episodes of constant star formation over two assembly periods, $t_2 - t_1$ and $t_1 - t_0$, which are defined by the halo growth times from $M_{h}/4$ to $M_{h}/2$ to $M_{h}$ respectively:
\begin{equation}
SFR(t_{i}, t_{i+1}, M_{h}) = \dfrac{\varepsilon(M_{h}/2^{i})M_{h}}{2^{i} (t_{i+1} - t_{i})}.
\label{eqn:sfr}
\end{equation}
The time $t_{i>0}$, given a halo of mass $M_h/2^{i-1}$ at $t_{i-1}$, is defined as the lookback time at which the progenitor of that halo is expected to have assembled a mass of $M_h/2^i$ based on the halo assembly time formalism of \citet{1993MNRAS.262..627L}. Additionally, $t_0$ is defined as the lookback time of the observed halo associated to the redshift $z$.

To derive the stellar efficiency, $\varepsilon(M_{h})$, we assume that the halos assemble by a median assembly time calculated with the ellipsoidal collapse model as given by \citet{2007MNRAS.376..977G}. For our stellar population model, we adopt a \citet{2003MNRAS.344.1000B} SSP with constant stellar metallicity $Z = 0.02~Z_{\odot}$ at all redshifts, and a Salpeter mass function ranging between $0.1~\mathrm{M_{\odot}}$ and $100~\mathrm{M_{\odot}}$. Given that high redshift observations from the WFC3's F160W band typically corresponds to UV light in the object's rest frame, we then define $l(a)$ as the luminosity at $1500\AA$ of an SSP with mass $1 M_{\odot}$ with an age $a$. The resulting luminosity of the halo is

\begin{equation}
\begin{aligned}
L(M_{h},z) 
= & \int^{t_1}_{t_0} SFR(t_1,t_0,M_{h})l(t-t_{0})dt \\
+ & \int^{t_2}_{t_1} SFR(t_2,t_1,M_{h})l(t-t_{0})dt.
\end{aligned}
\label{eqn:reslum}
\end{equation}

We calibrate $L(M_{h},z_{\mathrm{calib}})$ by abundance matching \citep{1999MNRAS.304..175M} the \citet{2013MNRAS.433.1230W} Universal Halo Mass Function (HMF) to the extrapolated LF from \citet{2015ApJ...803...34B} over a halo mass range of $10^{7} - 10^{14}M_{\odot}$ at $z_{\mathrm{calib}} = 4.9$. To ensure self-consistency, the Watson HMF parameters used are the $\tt{Tiamat}$ best fit parameters from \citet{2016MNRAS.459.3025P} and the same cosmological parameters of $\tt{Tiamat}$. Through this calibration, we use Eqs.~\ref{eqn:sfr} and \ref{eqn:reslum} to compute the redshift-independent $\varepsilon(M_{h})$. Eq.~\ref{eqn:reslum} can then be applied to obtain the corresponding galaxy luminosity against halo mass relation $L(M_{h},z)$ at any redshift (see \citealt{2015ApJ...813...21M} for more details on the model). 

\subsection{$k$ Shift Correction}

\begin{figure}[ht!]
	\centerline{\includegraphics[trim={0.6cm 0.6cm 0 0.3cm}, clip, angle=-00, scale=0.84]{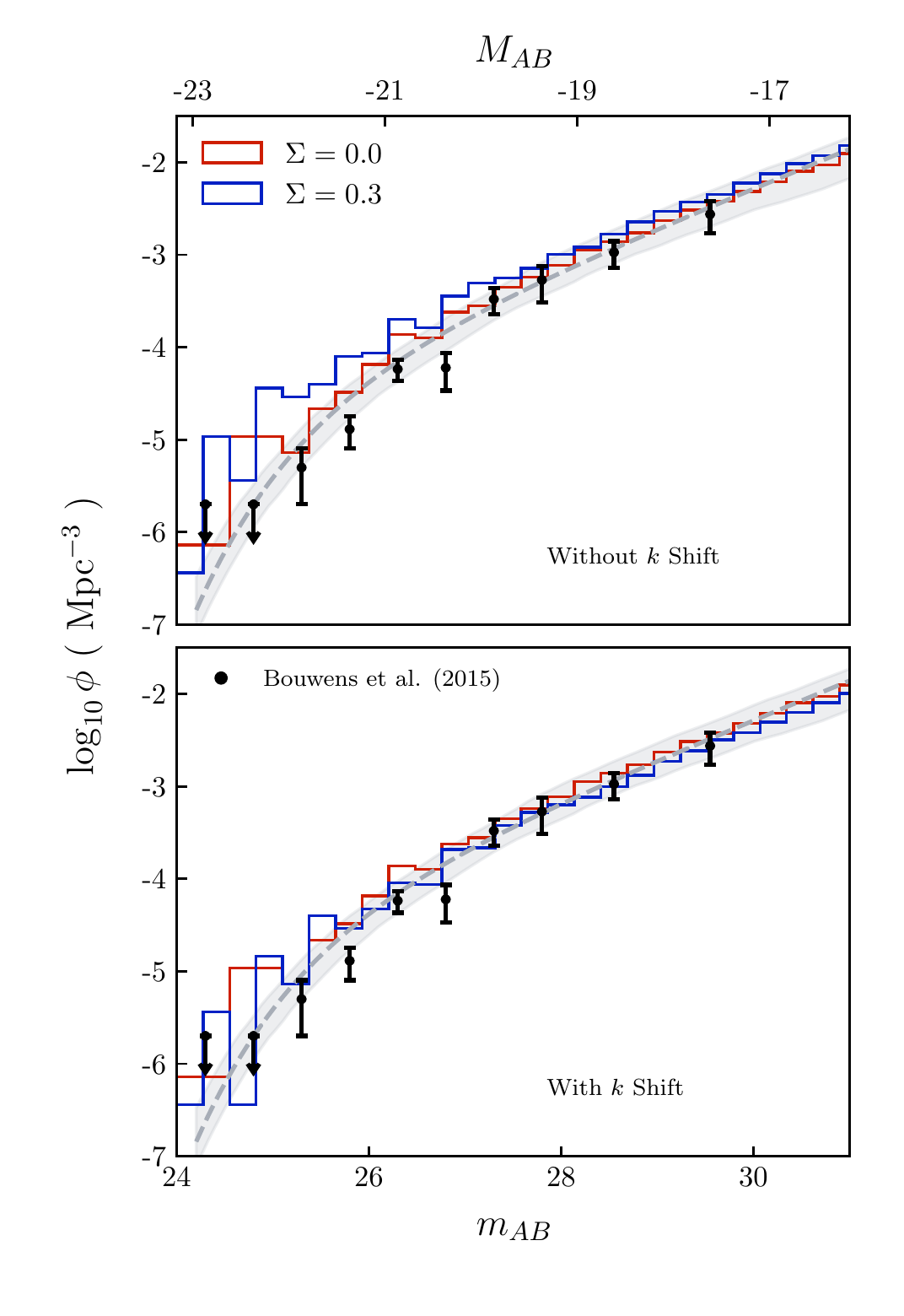}}
	\caption{\small LFs at redshift $z=7.86$ with $\Sigma$ values of $0, 0.3$. The LF for $\Sigma=0.3$ is derived from a single run after assigning galaxies to haloes. The grey dashed line is model UV LF with $\Sigma = 0$ and $1\sigma$ uncertainties. The black points are the observed LF values from \citet{2015ApJ...803...34B}. (Upper panel) No shift to the average galaxy luminosity, $k=0$. (Lower panel) $k=0.64$ shift to impose abundance matching of the LF at $m_{AB} = 26.2$.}
	\label{fig:complum}
\end{figure}

The $L(M_{h},z)$ model from \citet{2015ApJ...813...21M} is constructed assuming a one-to-one relation between halo mass and galaxy luminosity (i.e. $\Sigma=0$ in Eq~\ref{eqn:clf}). If we were to simply use $\langle L^{*}(M_{h}, \Sigma,z)\rangle \equiv L(M_{h},z)$, then we would introduce a general over-brightening of the luminosity function bright end for positive values of $\Sigma$, which is illustrated in the upper panel of Figure \ref{fig:complum} for $\Sigma=0.3$ (this effect becomes increasingly prominent with increasing $\Sigma$). Thus, to maintain a close fit to the observed LF, we add a first-order correction term, $k$, dependent on $z$ and $\Sigma$, to shift the mean galaxy luminosity in order to compensate for the LF brightening introduced by the scatter: 

\begin{equation}
\langle L^{*}(M_{h}, z, \Sigma)\rangle= L(M_{h}, z) - k(\Sigma,z). 
\label{eqn:k}
\end{equation}

The value of the parameter $k(\Sigma,z)$ is determined through abundance matching of the galaxy counts at a fixed luminosity, which here we set at $m_{AB} = 26.2$. The functional shape of $k$ is expected to be a monotonically increasing function in $\Sigma$ for a fixed $z$. Larger values of $\Sigma$ are expected to provide an increasingly exaggerated `bloat' compared to the $\Sigma = 0$ LF, and thus necessitate a larger correction. We iteratively determine the value of $k(\Sigma,z)$ by generating Monte Carlo realizations of galaxy catalogues with input dispersion $\Sigma$ at redshift $z$, measuring the average number density of objects with $m_{AB} < 26.2$ and calculating the average amount of offset, $k(\Sigma,z)$ required to match the number density of $m_{AB} < 26.2$ objects at $\Sigma = 0$.

Figure \ref{fig:complum} not only shows the effect of $\Sigma$ at the bright-end, but also illustrate visually that a log-normal dispersion preserves the power-law exponent at the faint end of the LF, as noted by \citet{2005ApJ...627L..89C}. From the lower panel of the figure, we can see how our first-order $k$-shift correction re-establishes a LF shape (blue line) that is very close to that derived without scatter (red), and an excellent model for the observed LF (black points with error-bars). 

\section{SCATTER IN LUMINOSITY VERSUS HALO MASS FROM THE HALO ASSEMBLY TIME DISTRIBUTION} \label{sec:cons}

An estimate of the lower bound of $\Sigma$ can be constrained from the distribution in halo assembly times, $t_{i}$. In fact, through Equation~\ref{eqn:reslum} a different halo assembly period, $t_{i+1} - t_{i}$ implies a different SFR and UV luminosity for the same halo mass. Thus, we can use the full probability distribution of the halo assembly time to derive the corresponding distribution of galaxy luminosity at fixed halo mass. For this, we start from the universal form of the distribution of halo assembly times for a halo with mass $M_{h}$ at redshift $z$ \citep{2007MNRAS.376..977G}: 

\begin{equation}
p(\omega) d\omega = 2 \omega \, \mathrm{ erfc}(\omega/\sqrt{2})d\omega, \\
\label{eqn:podo}
\end{equation}

where $\omega$ is a time-associated variable:

\begin{equation}
\omega = \sqrt{q}\dfrac{\delta_{c}(t_{1}) - \delta_{c}(t_{0})}{\sqrt{S(M_{h}/2) - S(M_{h})}}. \\
\label{eqn:omega}
\end{equation}

Here, $q = 0.707$ is a constant provided by \citet{2007MNRAS.376..977G} as a reasonable fit to an ellipsoidal collapse scenario. In addition, we have the critical density threshold, $\delta_{c}(t) = \delta_{sc}/D(t)$ where $\delta_{sc} \approx 1.69$ is the required linear density contrast for an overdensity to undergo spherical collapse, and $D(t)$ is the linear growth factor, appropriately normalized to unity at the present time (e.g. refer to \citealt{1992A&A...263...23B} for an analytic expression for the growth factor). Technically, Eqs.~\ref{eqn:podo} and \ref{eqn:omega} are derived under the special case of a white noise power spectrum for the initial density fluctuations, i.e. a scalar spectral index of $n=0$. However, the result is still an adequate approximation for our purposes as the distribution is relatively insensitive to $n$ \citep{1993MNRAS.262..627L, 2007MNRAS.376..977G}. 

\begin{figure}[ht!]
	\centerline{\includegraphics[angle=-00, scale=0.8]{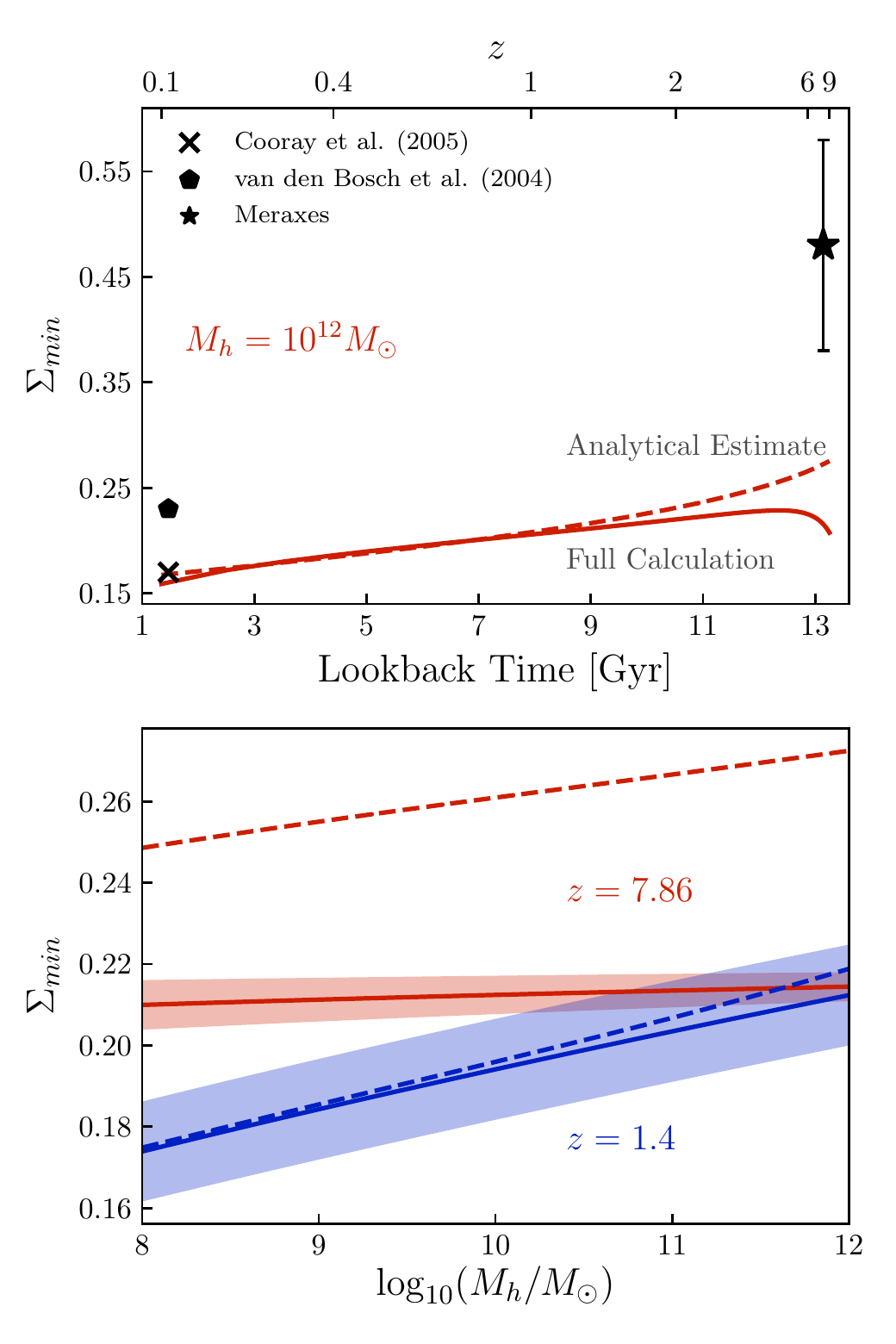}}
	\caption{\small (Upper panel) Evolution of $\Sigma_{min}$ over $z$ for a halo mass of $M_{h} = 10^{12}M_{\odot}$. (Lower panel) $\Sigma_{min}$ over $M_{h}$ at selected redshifts (colored curves). The solid line is $\Sigma$ derived with a Monte Carlo scheme sampling the halo assembly times and using Eq.~\ref{eqn:reslum}. The shaded region represents the $1\sigma$ confidence range. The dashed line is an analytical estimate as given by Eq.~\ref{eqn:approxsig}, which is a fair representation out to $z\sim2$ (upper panel), and a moderate overestimate at high $z$.}
	\label{fig:zvsSigma}
\end{figure}

In general, the calculation of the probability distribution of the galaxy luminosity has to be evaluated numerically, e.g. by sampling Eq.~\ref{eqn:podo} and then integrating numerically Eq.~\ref{eqn:reslum}, with results for the resulting scatter $\Sigma_{min}$ shown in Figure~\ref{fig:zvsSigma}. However, a convenient analytical solution for $\Sigma_{min}$ is available by noting that within our framework and to first order, the UV luminosity can be written as $L \propto 1/(t_{1}-t_{0})$, which is derived by assuming that our SSP has converged much earlier than the typical assembly period of the halo. From this we can substitute $L$ into Eq.~\ref{eqn:podo}: after transforming to a log scale, $\kappa = \log_{10}(L)$, the probability distribution of luminosity is

\begin{equation}
\begin{aligned}
p(\kappa) = \dfrac{2 A q \ln(10) \Phi}{10^{\kappa}} \Bigg[\dfrac{\delta_{c}(A/10^{\kappa} + t_{0}) - \delta_{c}(t_{0})}{S(M_{h}/2) - S(M_{h})} \Bigg] \\
\times \mathrm{ erfc}\Bigg(\sqrt{\dfrac{q}{2}} \dfrac{ \Big[\delta_{c}(A/10^{\kappa}+t_{0}) - \delta_{c}(t_{0})\Big]}{\sqrt{S(M_{h}/2) - S(M_{h})}}\Bigg)
\Bigg(\dfrac{d\delta_{c}(A/10^{\kappa} + t_{0})}{dt}\Bigg),
\end{aligned}
\end{equation}
\\
and the corresponding spread $\Sigma_{min}$

\begin{equation}
\Sigma_{min}^{2} = \Bigg(\int \limits^{\infty}_{\log(\frac{A}{T -  t_{0}})} \kappa^{2} p(\kappa)d\kappa \Bigg) - \Bigg(\int \limits^{\infty}_{\log(\frac{A}{T -  t_{0}})} \kappa p(\kappa)d\kappa\Bigg)^{2} \\,
\label{eqn:approxsig}
\end{equation}

where $T$ is the age of the universe, $\Phi$ is a normalization constant and $A$ is a proportionality constant relating $1/(t_{1}-t_{0})$ to $L$. The log-normal scatter, $\Sigma_{min}$ is insensitive to $A$, as the factor only provides an offset to the $\log(L)$ versus $t_{1}-t_{0}$ relation. $\Sigma_{min}$ calculated this way depends on the halo mass and redshift. Figure \ref{fig:zvsSigma} illustrates the results obtained exploring a range of the input parameters, where the dashed lines are $\Sigma_{min}$ derived through Eq.~\ref{eqn:approxsig} and the solid line is $\Sigma_{min}$ from the detailed Monte Carlo sampling of the halo assembly time distribution. In both cases, the calculated scatter have the trend of a rapidly increasing $\Sigma_{min}$ up to $z\sim 2-3$ and stabilizing at higher $z$. Differences in $\Sigma$ between methods at higher $z$ is largely attributed to the shortening of the assembly period, into the regime where the SSP still experiences notable change in its total luminosity over cosmic time. We also note that the full Monte Carlo calculation is essentially insensitive to halo mass at high redshift (lower panel of Fig \ref{fig:zvsSigma}), while the analytical approximation gives a weak scaling of $\Sigma_{min}$ with $M_{h}$ at a fixed $z$.

As expected, at low redshifts ($z \lesssim 2$), the full Monte Carlo measurement of $\Sigma$ is well approximated by Eq.~\ref{eqn:approxsig}, and we infer a minimum $\Sigma_{min} \sim 0.17$ at $z\sim 0$ for halos with $M_{h} = 10^{12} ~\mathrm{M_{\odot}}$. At high redshifts ($z \gtrsim 4$), the estimate of $\Sigma_{min}$ from that equation is still within $\lesssim 20\%$, and we find overall $\Sigma_{min} \sim 0.2$. 

\begin{figure}[ht!]
	\centerline{\includegraphics[angle=-00, scale=0.8]{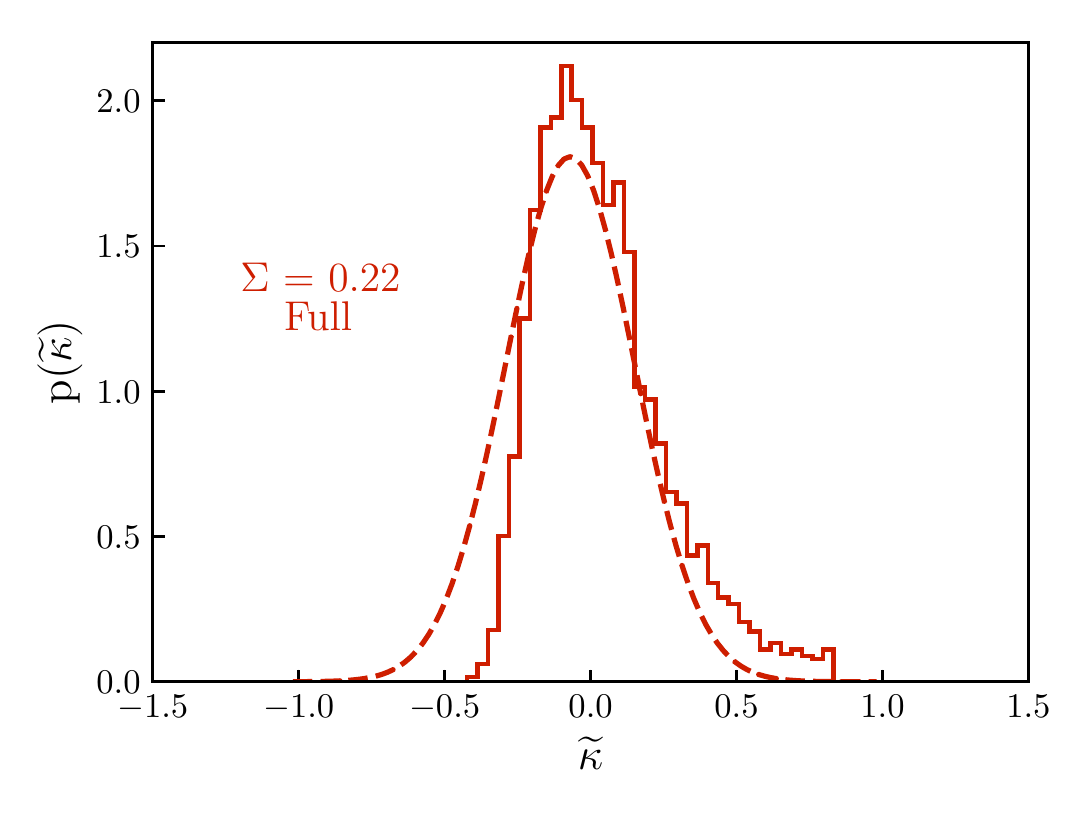}}
	\caption{\small Probability distribution of $\widetilde{\kappa} = \kappa - \langle \kappa \rangle$. This plot was generated at $z=7.86$ with $M_{h} = 10^{10} M_{\odot}$. The dashed line is the log-normal curve overlayed on the distribution of $\kappa$ derived from Equation~\ref{eqn:reslum} via Monte Carlo sampling (solid line).}
	\label{fig:lognormalScatter}
\end{figure}

In Figure~\ref{fig:lognormalScatter} we see the shape of the underlying distribution of $\kappa$ (offset with $\widetilde{\kappa} = \kappa - \langle \kappa \rangle$), generated from the Monte Carlo sampling and overlaid with the log-normal equivalent using the calculated value of dispersion. There is a clear skew towards higher luminosities, originating from the skewed nature of the distribution in halo assembly times. The analytical approximation of Equation~\ref{eqn:approxsig} shows a similar characteristic shape, justifying the assumption of a log-normal scatter to first order. One caveat is that if we were to include other possible sources for scatter in luminosities at fixed halo mass, the relative distributions might have different shapes, hence it is difficult to predict with confidence the overall functional form of the probability distribution for $\kappa$.

Additionally, the estimates for $\Sigma_{min}$ shown in Figure~\ref{fig:zvsSigma} can be compared against observational measurements of $\Sigma$ at low redshift. \citet{2005ApJ...627L..89C} find $\Sigma \sim 0.23$ best fits the LF from the sample of \citet{2003ApJ...584..203H} ($z_{median} \sim 0.14$ and a characteristic halo mass, $M_{h} = 10^{12.6}M_{\odot}$), which is higher than our lower limit of $\Sigma_{min} \sim 0.18$ for the same redshift and halo mass. From \citet{2004bdmh.confE..41V}, we can extract $\Sigma \sim 0.17$ from the galaxy $L$ vs $M_{h}$ relation using the 2dFGRS data from \citet{2002MNRAS.333..133M} at a similar redshift $z_{median} \sim 0.11$, and this measurement is closer to, but still above, our minimum scatter limit. At higher redshift there are no empirical measurements of $\Sigma$, but we can still compare our results against more detailed cosmological simulations of galaxy formation. For this, we analyzed the semi-analytic galaxy model of Meraxes \citep{2016MNRAS.462..250M}, implemented over the same cosmological dark-matter only simulation considered here. We measure $\Sigma_{\mathrm{Meraxes}} \sim 0.38 - 0.58$ at $z=7.86$, where $\Sigma_{\mathrm{Meraxes}}$ is dependent on the considered halo mass range. The lower limit of $\Sigma_{\mathrm{Meraxes}} \sim 0.38$ is attained by considering all halos with mass $M_{h} > 10^{10.7} M_{\odot}$ (hosting central galaxies with $m_{AB} \lesssim 28$), and the upper limit $\Sigma_{\mathrm{Meraxes}} \sim 0.58$ is through considering all resolved halos in the simulation with $M_{h} > 10^{9} M_{\odot}$. The $\Sigma_{\mathrm{Meraxes}}$ extracted from Meraxes is substantially larger, but still consistent with the lower limit $\Sigma_{min} \sim 0.25$ from our modeling.

Thus, to build upon a fully cohesive interpretation, consistent with Meraxes, we suggest the contribution of an an additional source of scatter from the stochasticity in $\varepsilon (M_h)$, with a small trend to increase for lower halo masses. From the differences in $\Sigma$ in low redshift studies, we can infer up to $\Sigma_{\varepsilon}\sim 0.2$, and up to $\Sigma_{\varepsilon} \sim 0.5$ from the scatter extracted in Meraxes. Under this scenario, since halo assembly time and star formation efficiency are independent, their contributions to $\Sigma$ are summed in quadrature.

In principle, galaxy to galaxy variations in dust content can also contribute to the amount of scatter in UV luminosities. In this case, the overall $\Sigma$ would include an additional term $\Sigma_{\mathrm{dust}}$, to be summed in quadrature with $\Sigma_{min}$ and $\Sigma_{\varepsilon}$ to derive the total scatter. 

Motivated by this analysis, we primarily focus on models with $0.3\leq \Sigma \leq 0.5$ for this work. Still, we include $\Sigma = 0.1$ as a comparison case for a (unphysical) small scatter value. 

\section{RESULTS AND DISCUSSION}  \label{sec:results}

\begin{figure}[ht!]
	\centerline{\includegraphics[trim={0cm 1.2cm 0.5cm 2.2cm}, clip, angle=-00, scale=0.8]{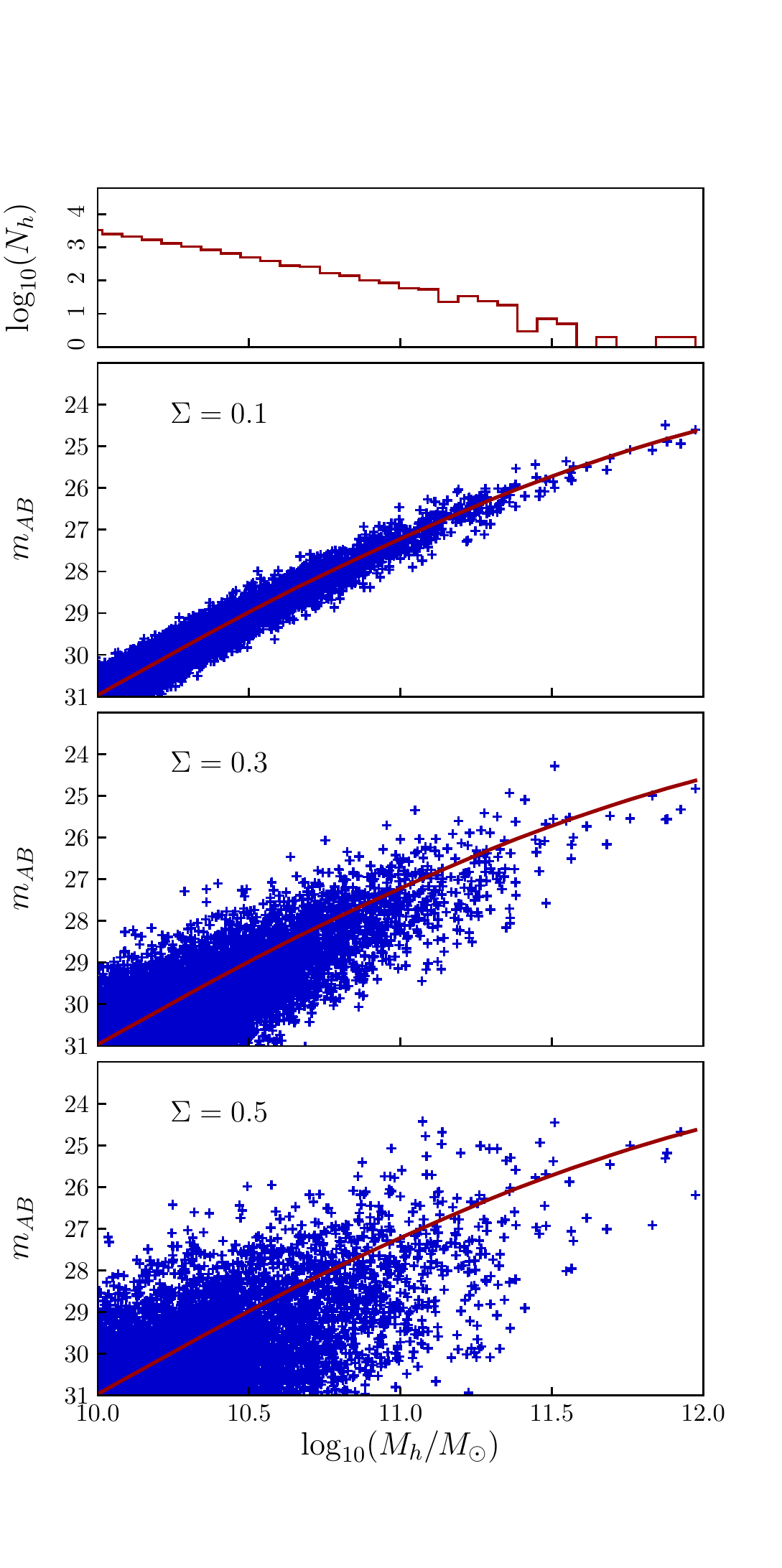}}
	\caption{\small A single Monte Carlo realization of galaxies inside the simulation volume at $z=7.86$. (First panel) The number of dark matter haloes at each mass bin. (Second panel) Haloes populated with galaxies with a dispersion factor, $\Sigma = 0.1$. The most luminous galaxy at $m_{AB} \approx 24.5$ is found within a halo of mass, $M_{h} \approx 10^{11.8}M_{\odot}$. (Third panel) The second panel, but with $\Sigma = 0.3$. The most luminous galaxy at $m_{AB} \approx 24.2$ is found within a halo with mass, $M_{h} \approx 10^{11.5}M_{\odot}$. (Fourth panel) The second panel, but an extreme case with $\Sigma =0.5$. Here the most luminous galaxy is $m_{AB} \approx 24.2$ in a halo with mass, $M_{h} \approx 10^{11.1}M_{\odot}$.  In all scatterplots, $L(M_{h})$ is given by the solid line. Additionally, note that the galaxies aren't symmetrically distributed around $L(M_{h})$ because of first-order $k$ shift correction.}
	\label{fig:sigmaillu}
\end{figure}

With the setup described in Section~\ref{sec:method}, we ran $2.5 \times 10^{4}$ realizations of galaxy catalogs and pencil beams through the catalogs for each configuration considered, grouped into 10 sets of $2.5 \times 10^{3}$ realizations to determine a characteristic $2\sigma$ uncertainty between realizations via bootstrapping. The results from a single realization are illustrated in Figure~\ref{fig:sigmaillu} for different values of $\Sigma$. The figure illustrates well the general impact of increasing the scatter: Because there are many more low-mass halos compared to rarer higher mass halos, increasing $\Sigma$ leads to having a higher likelihood that the most luminous object is not hosted in the highest mass halo. In turn, this would affect the clustering strength around the brightest simulated galaxies. 
 
\begin{figure*}[b!]
\centerline{\includegraphics[angle=-00, scale=0.56]{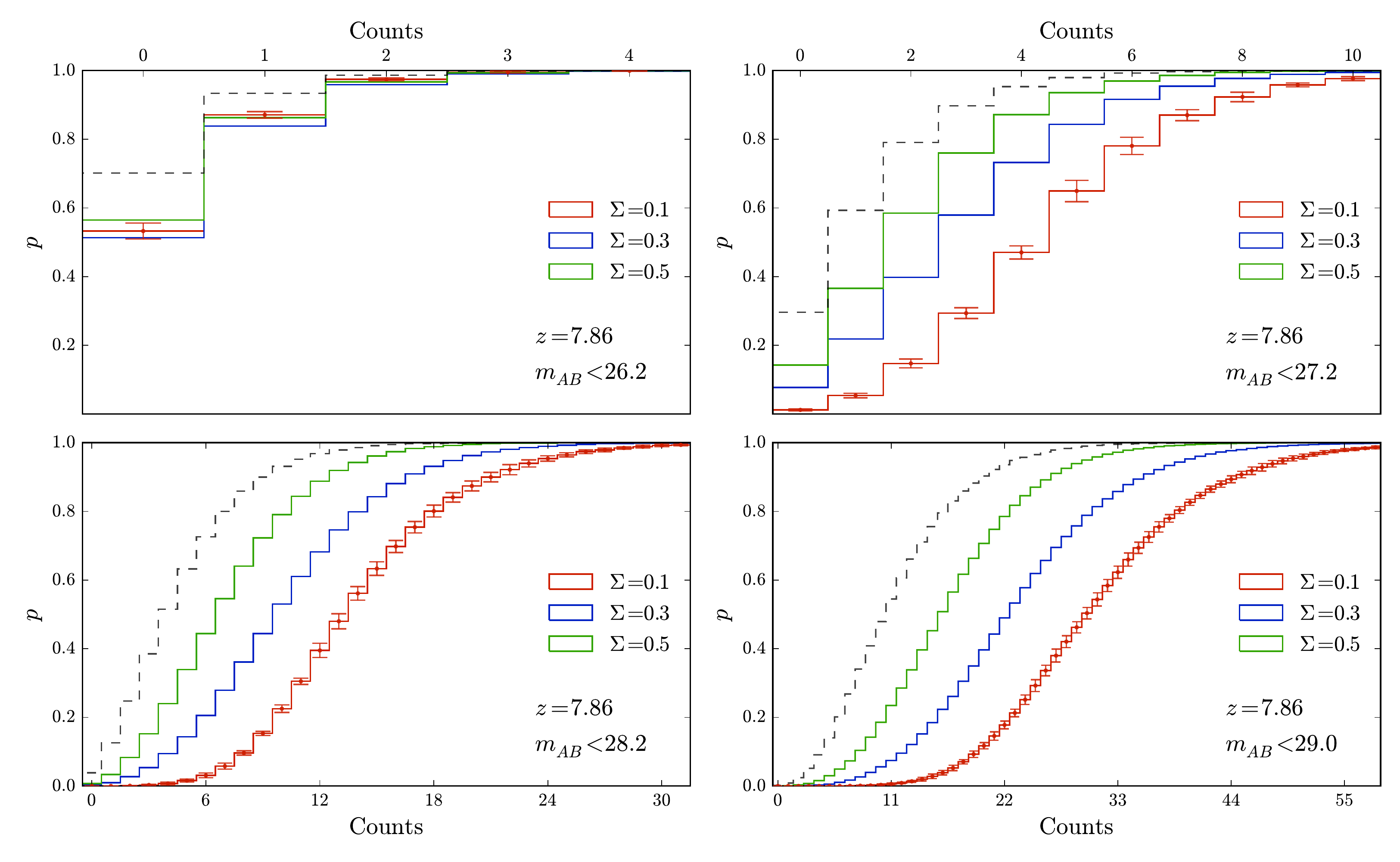}}
\caption{\small Cumulative distribution function for predicted counts of neighbor galaxies in a WFC3 field-of-view centered around a $m_{AB}\sim 25$ galaxy at $z=7.86$ for different values of the scatter parameter ($\Sigma$ - colored lines) and with different panels illustrating different magnitude limits:  (Upper left) $m_{AB} < 26.2$, (Upper right) $m_{AB} < 27.2$, (Lower left) $m_{AB}<28.2$, (Lower Right) $m_{AB}<29.0$. A neighbor is classified as any galaxy brighter than the imposed magnitude limit, and excluding the central target source. For reference, the dashed line shows the predicted counts for blank field (random pointing) for the model with $\Sigma = 0.5$. 90\% confidence error bars are shown as reference for $\Sigma=0.1$.}
\label{fig:z8-26}
\end{figure*}

\begin{figure*}[t]
\centerline{\includegraphics[angle=-00, scale=0.56]{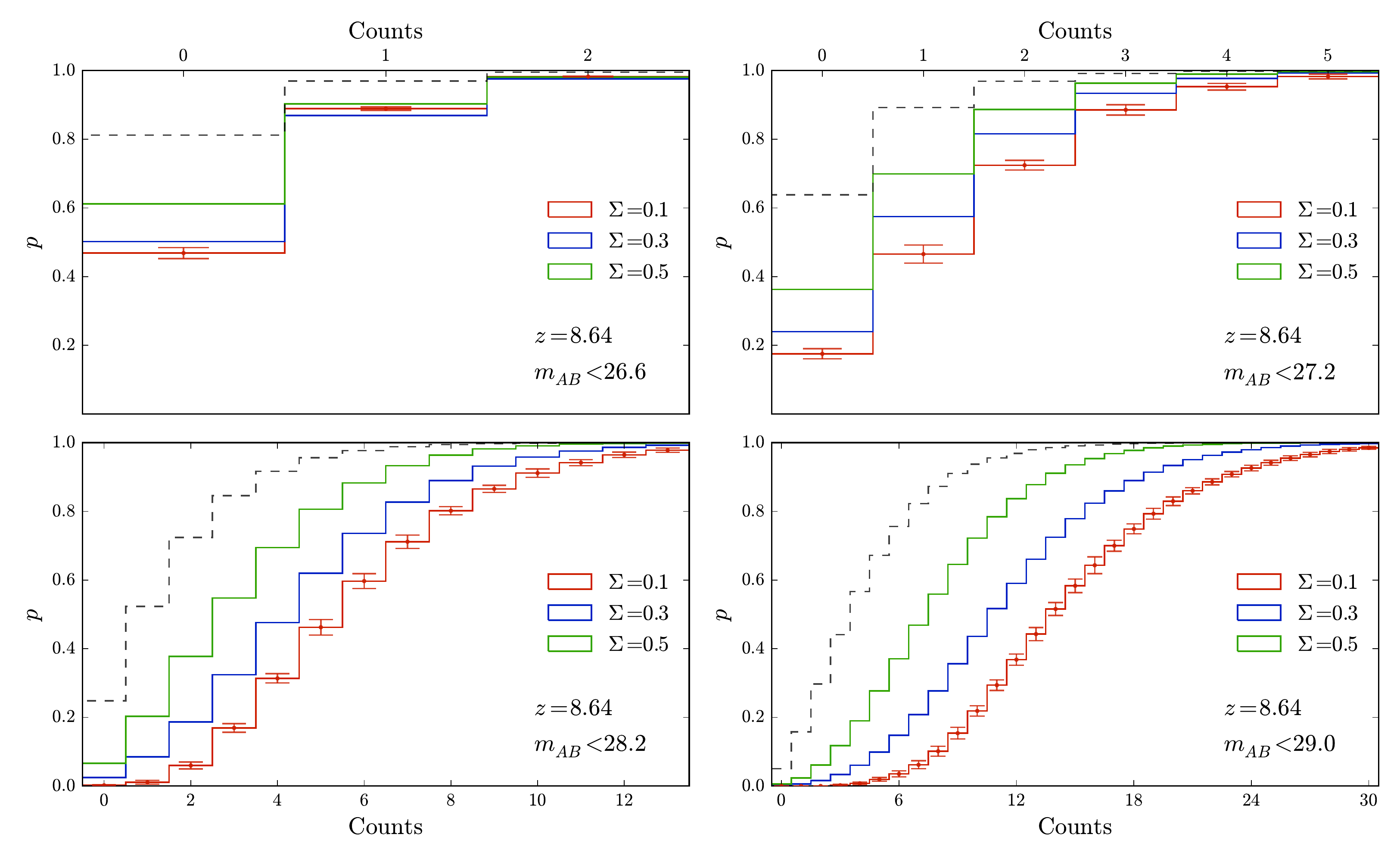}}
\caption{\small As in Figure~\ref{fig:z8-26} but for pencil beams centered at $z = 8.64$. \vspace{0.265cm}}
\label{fig:z9}
\end{figure*}

Figures \ref{fig:z8-26} and \ref{fig:z9} show the full results from the simulations, reporting the predicted cumulative number count distribution of galaxies visible inside pencil beams at two different redshifts, $z=7.86$ and $z=8.64$ and under 2 scenarios: 1) the pencil beam is centered around the most luminous object of the catalog, and 2) the pencil beam is placed randomly inside the simulation volume. We vary $\Sigma$ in Eq.~\ref{eqn:clf} to quantify the changes in clustering. Galaxies are counted up to a given limiting magnitude, and the counts exclude the brightest central object for targeted pencil beams. From the figures it is clear that higher $\Sigma$ values decrease the probability of having large number counts in regions surrounding the brightest galaxies, and the fainter the limiting magnitude, the cleaner is the separation between curves at different $\Sigma$. The first trend can be understood by considering that increasing $\Sigma$ implies an increase in the odds that the brightest galaxy in the realization is hosted into a lower mass halo. This arises because increasing the scatter in the luminosity versus halo mass relation may produce a lower-mass over-luminous outlier, facilitated by the steep halo mass function (see Fig.~\ref{fig:sigmaillu}). The second trend (better discrimination with deeper observations) derives from a lesser impact of Poisson noise when average counts are higher. 

\begin{deluxetable}{llrrr}[t!!]
\tablewidth{0pt}
\tablecaption{Probabilities of observing zero neighbors in specified magnitude range}
\tablehead{\colhead{Redshift} & \colhead{$\Sigma$} & \colhead{$p$(0 neighbors)} & \colhead{$p$(0 neighbors)} & \colhead{$p$(0 neighbors) \label{tab:beamparam}} \\ 
\colhead{} & \colhead{} & \colhead{$m_{AB} < 26.2$} & \colhead{$m_{AB} < 26.6$} & \colhead{$m_{AB} < 27.2$}  } 
\startdata
$z = 7.86$	&	0.1	&	$0.53 \pm 0.023$	&	$0.12 \pm 0.012$		&	$0.012 \pm 0.003$	\\
			&	0.3	&	$0.51 \pm 0.028$	&	$0.29 \pm 0.018$		&	$0.077 \pm 0.010$	\\ 
			&	0.5	&	$0.56 \pm 0.027$	&	$0.37 \pm 0.019$		&	$0.14 \pm 0.009$	\\
			&		&					&						&					\\ 
$z = 8.64$	&	0.1	&	$0.60 \pm	0.016$	&	$0.47 \pm 0.019$		&	$0.18 \pm	0.015$	\\
			&	0.3	&	$0.67 \pm	0.013$	&	$0.50 \pm 0.025$		&	$0.24 \pm	0.017$	\\
			&	0.5	&	$0.75 \pm	0.015$	&	$0.61 \pm 0.018$		&	$0.36 \pm	0.019$	
\enddata
\tablecomments{A neighbor is classified as a galaxy detected inside the pencil beam that is not the target galaxy. The $2\sigma$ uncertainty is included with the probabilities.}
\label{tab:prob26}
\end{deluxetable}

Results from the simulations are also summarized in Table \ref{tab:prob26}. Here, we focus on the probability of observing zero neighbors inside our pencil beam when targeting brightest galaxy at three magnitude limits, $m_{AB} < 26.2$, $ 26.6$ and $ 27.2$. The first magnitude limit was selected to be consistent with the expected completeness limit of the dropout galaxy search for the CANDELS-EGS survey (the z-band filter has a limiting magnitude $m_{AB} =26.1$, which represents the bottle neck in identification of dropout sources). The other cases investigate the impact of deeper limiting magnitudes, and forecast neighbor count probabilities for future surveys. 

At the shallowest depth, we note that the probability of observing $0$ neighbors is dominated by noise due to the low object counts at the lower magnitude limits ($m_{AB} < 26.2$ for $z = 7.86$ and up to $m_{AB} < 26.6$ for $z=8.64$). For $z = 7.86$, typical number densities of $M_{AB} \approx -21$ galaxies are of the order of $\sim 10^{-5} \mathrm{Mpc}^{-3}$ comoving. Thus, we can expect $\sim 10^{1}$ objects brighter than $m_{AB} < 26.2$ in our simulation volume of $100$ $\mathrm{Mpc}^{3}$. The low-number statistics is even more extreme for the second redshift case ($z=8.64$; fig~\ref{fig:z9}) as number counts for $m_{AB} < 26.2$ (or even $m_{AB}< 26.6$) galaxies fall to the order of $\sim 10^{0}$. These finding imply that to the magnitude limit of the EGS field ($m_{AB}<26.2$), no clustering signal is expected to be detected through measurement of an overdensity of neighbors in a single WFC3 field centered around a $L>L_*$ source at high $z$. Interestingly, by reaching a deeper sensitivity ($m_{AB}<26.6$) for $z=7.86$, the detection of an overdensity becomes statistically more significant ($\sim 90\%$ confidence of identifying at least one neighbor) for the minimum scatter case, and should still be relatively likely (down to 60\% confidence) for the range of scatter values ($0.1<\Sigma<0.5)$. For $m_{AB}<27.2$ the detection of at least one neighbor becomes highly likely ($\sim 99\%$ confidence), and it starts to become possible to discriminate different values of $\Sigma$ from the cumulative distribution function (see top right panel of Fig.~\ref{fig:z8-26}), which could be measured observationally by a hypothetical follow-up of the brightest sources identified by Hubble at $z\sim 8$. The situation at $z=8.64$ is qualitatively similar, but since sources are rarer, the impact of Poisson noise is more pronounced, and observations reaching about $0.5$ mag deeper are needed to achieve a similar degree of inference on clustering. 

The trends we identified are consistent with the measurements of clustering through the two point correlation function, as reported for example by  \citet{2014ApJ...793...17B} at $z\sim 4-8$ using Hubble observations, and more recently at $z = 4 \sim 7$ from the ground \citep{2017arXiv170407455O, 2017arXiv170406535H}. In fact, luminosity-dependent clustering is measured only for samples of objects with a relatively high number density ($z\lesssim 5$ for Hubble data) or by leveraging observations over a wide area (as in the latter case thanks to Subaru's Hyper Suprime Camera large field of view). 

The motivation for this study was to understand whether the lack of neighbors around EGS-zs8-1 and EGSY8p7 at similar redshift was creating tension with modeling of galaxy formation and evolution. Our results provide a natural explanation for the absence of a clustering signature in the data at the current depth. In addition, this framework also provides a theoretical/numerical interpretation for the clustering around one protocluster candidate at $z\sim 8$ from the Hubble BoRG survey \citep{2012ApJ...746...55T}. Follow-up imaging reported in \citet{2014ApJ...786...57S} identified the presence only a weak overdensity of photometric candidates, which can be explained by assuming $\Sigma \gtrsim 0.2$. While these findings highlight that current observations are in general not sufficiently deep to clearly detect over-densities around the brightest galaxies during the epoch of reionization, the future looks very promising. In fact, thanks to the James Webb Space Telescope, it will be possible to investigate protocluster environments to $m_{AB}\sim 28.2$ with just one hour of observing time (at $S/N \sim 5$ with four broad-band filters), unveiling the nature of over-densities around bright sources (Figs.~\ref{fig:z8-26}-\ref{fig:z9}. This will allow us to constrain the value of $\Sigma$ observationally, which in turn can be used to falsify predictions from models of galaxy formation.

\section{CONCLUSION} \label{sec:disc}

In this work we extend the physically motivated galaxy luminosity function modeling framework of \citet{2010ApJ...714L.202T}, \citet{2013ApJ...768L..37T} and \citet{2015ApJ...813...21M} to include for stochasticity in the luminosity at fixed dark-matter halo mass and investigate galaxy clustering observations around rare $L>L_*$ sources at high redshift. Specifically, the innovation lies in introducing a scatter in galaxy luminosities that is modeled as a log-normal parameter, $\Sigma$, using a conditional luminosity function (CLF) approach, which is a well established tool to investigate the relation between luminosity and halo mass at lower redshifts \citep{2003MNRAS.339.1057Y, 2004MNRAS.353..189V, 2005ApJ...627L..89C}. The analytical CLF is then combined with a high-resolution dark-matter only cosmological simulation \citep{2016MNRAS.459.3025P} to construct Monte Carlo mock catalogs of high-redshift sources for both random pointings and for lines of sight centered toward the most luminous sources in a given volume. The values of $\Sigma$ considered in our Monte Carlo experiments include a range with a lower limit set by the contribution to luminosity scatter deriving from the halo assembly time distribution, and an upper limit broadly motivated by comparison with observations and cosmological simulations (Section~\ref{sec:cons}). Our key findings can be summarized as follows:

\begin{itemize}

\item By including a scatter in the $L(M_h)$ relation, we are naturally breaking the tight relation between most luminous galaxy and most massive halo. This is exemplified in Figure~\ref{fig:sigmaillu} which highlights the possibility for an over-luminous galaxy in a smaller-mass halo to be brighter than all sources hosted in more massive halos. The larger $\Sigma$, the more likely and significant this scenario, especially because of the steep dark matter halo mass function at the high-mass end, which boosts the chance that one of the many common lower mass halos outshines all their more massive rarer peers. 

\item The halo assembly time distribution sets a lower bound $\Sigma_{min}$ in our framework (Eq.~\ref{eqn:approxsig}), with results shown in Figure~\ref{fig:zvsSigma} illustrating that $0.1\lesssim \Sigma_{min} \lesssim 0.3$ increases with redshift. When compared to observational results of luminosity scatter in low redshift galaxies (Section~\ref{sec:cons}), we found that our minimum values are justified. In addition, the lower bound of $\Sigma_{min}$ is consistent with the scatter measured from the semi-analytical model, Meraxes. We also further account for the possibility of an additional contribution from $\Sigma_{\varepsilon}$. Overall, this motivates our decision to investigative the range $0.1 \leq \Sigma \leq 0.5$.

\item Monte Carlo mock observations of Hubble's WFC3 pointings ($\approx 4.5$ arcmin$^2$) by applying our CLF model over the DRAGONS dark matter halo catalogues, $\tt{Tiamat}$, and tracing pencil beams through the volume, with results shown in Figures~\ref{fig:z8-26} - \ref{fig:z9} and in Table~\ref{tab:prob26}. We find that while pencil beams centered on very bright galaxies show an excess of number counts compared to random lines of sight, the difference is challenging to measure in typical (shallow) surveys with Hubble, reaching only $m_{AB}\sim 26-27$. Thus, the lack of clustering signal around two bright spectroscopically confirmed sources at $z = 7.86$ and $z=8.64$ is consistent with our modeling. The main reason is that, at the magnitude limit of current observations, the average number of galaxies that are expected to be detected is low, hence Poisson noise is comparable to the signal. 

\item Deeper observations (reaching  $m_{AB}\gtrsim 27-28$ at $z\sim 8$ and $z\sim 9$ respectively) are however predicted to be able to detect a clustering signal in the environment of $L>L_*$ sources, and could in principle lead to an observational determination of $\Sigma$ if multiple fields are observed. This highlights the potential for future observations to investigate indirectly the stochasticity of halo-to-halo variations in star formation efficiency and assembly times, providing an innovative constraint on models of early galaxy formation and evolution. Such observations would be easily enabled by just one hour of observing time with JWST per pointing.
\end{itemize}

Finally, the qualitative results from this analysis can be applied more broadly to studies of galaxy clustering in extreme environments, including those that host the bright SDSS $z\sim 6$ quasars (QSOs; \citealt{2006AJ....132..117F}). The theoretical expectations has been that rare QSOs are hosted in rare, high-mass dark-matter halos \citep{2005Natur.435..629S, 2009MNRAS.394..577O,2011ApJ...736...66R, 2016A&ARv..24...14O}, but follow-up observations with Hubble have not detected a statistically significant overdensity of galaxies \citep{2009ApJ...695..809K, 2014AJ....148...73M}. The tension between modeling and observations can be alleviated or resolved by considering not only a scatter in the galaxy luminosity versus halo mass, but also especially in the QSO luminosity versus halo mass, which could be treated with a modeling framework similar to the approach pursued here for galaxies, demonstrating the value of constructing basic yet physically motivated models to connect dark to luminous matter to improve our understanding of the epoch of reionization. \\

\acknowledgements

We would like to thank the anonymous reviewer for their helpful comments and input, Charlotte Mason for helpful suggestions and for sharing her LF model code, and the DRAGONS team for sharing the DM halo catalogues. This research was conducted by the Australian Research Council Centre of Excellence for All Sky Astrophysics in 3 Dimensions (ASTRO 3D), through project number CE170100013. K.R is additionally supported through the Research Training Program Scholarship from the Australian Government. This work is based in part on Hubble Space Telescope projects HST GO-12905, 13767, and 15212.

\bibliographystyle{aasjournal}
\bibliography{clusterpaper}

\end{document}